\documentclass[a4paper,11pt]{article}
\usepackage{amsmath}
\usepackage{amssymb}
\usepackage{graphicx}
\begin{document}
\topmargin 0pt \oddsidemargin 0mm


\begin{center}
{\Large \bf Modified holographic dark energy in DGP brane world } \vspace{20mm}

{\large Dao-Jun Liu,  \footnote{e-mail address:
djliu@shnu.edu.cn}} {\large Hua Wang,   Bin Yang}

\vspace{1mm} {\em  Center for Astrophysics,
Shanghai Normal University, 100 Guilin Road, Shanghai
200234,China}

\end{center}
\date{ }

\begin{abstract}
In this paper,  the cosmological dynamics of a modified holographic dark energy which is derived from the UV/IR duality by considering the black hole mass in higher dimensions as UV cutoff, is investigated  in Dvali-Gabadaze-Porrati (DGP) brane world model. We choose Hubble horizon and future event horizon as IR cutoff respectively. And the two branches of the DGP model are both taken into account. When Hubble horizon is considered as IR cutoff, the modified holographic dark energy (HDE) behaves like an effect dark energy that modification of gravity in pure DGP brane world model acts and it can drive the expansion of the universe speed up at late time in $\epsilon=-1$ branch which in pure DGP model can not undergo an accelerating phase. When future event horizon acts as IR cutoff, the equation of state parameter of the modified HDE can cross the phantom divide.
\end{abstract}

PACS numbers: 98.80.-k, 95.36.+x

\section{Introduction}

Strong evidences from the current cosmological observations such as supernovae type Ia (SNeIa)\cite{A.G. Riess}, cosmic microwave background (CMB)\cite{D. N. Spergel et al.} and large scale structure (LSS)\cite{M. Tegmark et al.} converge upon the fact that the universe is spatially flat and there exists exotic component, dubbed dark energy, which drives the speed-up expansion of the universe. Many scenarios have been proposed to explain the acceleration. The preferred and simplest candidate for dark energy is the Einstein's cosmological constant which can fit the observations well. However, it suffers from the so-called fine-tuning problem  and coincidence problem.
The cosmological constant  corresponds to a perfect fluid with an equation of state parameter $w_{\Lambda}=-1$ constant in time and space. However, the current observations say little about the variation on time of such a parameter.
 Therefore, many dynamical models have also been studied extensively, such as quintessence \cite{Quintessence}, phantom\cite{Phantom}, quintom\cite{Quintom},  tachyon\cite{tacyon}, generalized Chaplygin gas\cite{cg}, etc (see \cite{copeland:2006ej} for a comprehensive review on dark energy models).

 The dark energy problem may be essentially an issue of quantum gravity \cite{Witten:2002}. Although a complete  theory of quantum gravity has not been established, some valuable ideas are thought to be the features of the theory of quantum gravity, for example, holographic principle and extra dimensions. Holographic principle says that the degrees of freedom in a bounded system should be finite and does not scale with its volume but with its boundary area \cite{G. t Hooft}. Besides, it is also suggested that the ultraviolet (UV) cutoff scale of a system is connected to its infrared (IR) cutoff scale. This relationship  is sometimes called UV/IR duality. Cohen et al\cite{A. Cohen} pointed out that, for a system with size $L$ and UV cutoff $\Lambda$ without decaying into a black hole, the quantum vacuum energy of the system should not exceed the mass of a black hole with the same size, so $L^3\rho_{\Lambda}\leq LM_p^2$, where $\rho_{\Lambda}$ is the vacuum energy density and $M_p=({{8\pi G}})^{-1/2}$ is reduced Plank mass.  Applying this idea to cosmology, one can choose the largest IR cutoff $L$ which saturates the inequality and take the dark energy density $\rho_{de}$ as $\rho_{\Lambda}$. Therefore, $\rho_{de} = 3c^2 M^2_pL^{-2}$,  where $c$ is a numerical constant. By choosing a reasonable IR cutoff scale, the density and equation of state is close to the observed value and, therefore, this kind of dark energy, which is usually called holographic dark energy (HDE), can be taken as  viable models\cite{Limiao}. The phenomenological consequences of these models have been studied extensively \cite{HDE}.

On the other hand, the  extra dimensions  exist in some fundamental theories,
especially, in string/M theory. To be more specific, as is well known, in brane world models,
where the spacetime dimension is more than four, black holes will in general be $D$-dimensional,
no matter what their $4$-dimensional effective effects could be. Therefore, it is interesting to investigate
holographic dark energy by applying the UV/IR duality to high dimensional theories, such as brane world models.
In Ref.{\cite{Saridakis}, Saridakis apply the bulk holographic dark energy in general $5$-dimensional two-brane models
and show that the equation of state of the effective $4$-dimensional holographic dark energy can cross
the phantom bound $w_{\Lambda}=-1$.

More recently, Gong and Li \cite{Gong:2010}, by using the mass of black holes in higher dimensional spacetime, proposed a modified holographic dark energy  model with Hubble scale as the IR cutoff and investigate this modified holographic dark energy in usual Friedmann universe. In this paper, we shall consider the cosmological evolution of the holographic dark energy from the  modified UV/IR duality in both branches of Dvali-Gabadaze-Porrati (DGP) brane model.

The rest of the paper is organized as follows. In section \ref{sec2}, we
present fundamental equations for the modified HDE in DGP brane world model, then analyze the model by choosing Hubble horizon as IR cutoff in section \ref{sec3}, and calculate the equation of state and decrease parameter in the case that future event horizon as IR cutoff in section \ref{sec4}.  In the end we draw some conclusions.


\section{Modified HDE and DGP brane world}
\label{sec2}
In $N+1$ dimensional space-time, the mass of the Schwarzschild black hole (SBH) is given by \cite{Myers:1987}
\begin{equation}
M_{SBH}=\frac{(N-1)A_{N-1}}{16\pi G}r_H^{N-2},
\end{equation}
where $A_N$ denotes the dimensionless area of a unit $N$-sphere, and $r_H$ is the horizon scale of the black hole. And $G$ is the $N+1$ dimensional gravitational constant, which is related to the $N+1$ dimensional Planck mass $M_{N+1}$ and the usual Planck mass $M_p$ in 4-dimensional space-time via
\begin{equation}
8\pi G=M_{N+1}^{-(N-1)}
\end{equation}
\begin{equation}
M_p^2=M_{N+1}^{N-1}V_{N-3},
\end{equation}
where $V_{N-3}$ is the volume of the extra-dimensional space.
Therefore, we have
\begin{equation}
M_{SBH}=\frac{(N-1)A_{N-1}}{2V_{N-3}}M_p^2r_H^{N-2}.
\end{equation}
Using the assumption of Ref.\cite{Gong:2010}, we have the relation
\begin{equation}
L^3\rho_{de}\sim\frac{(N-1)A_{N-1}}{2V_{N-3}}M_p^2L^{N-2}
\end{equation}
and, therefore, the modified holographic dark energy density reads
\begin{equation}
\rho_{de}=c^2\frac{(n+2)A_{n+2}}{2V_{n}}M_p^2L^{n-2},
\end{equation}
where $n=N-3$ is the number of extra dimensions and $c$ is a constant. For the case that there is only one extra dimension(i.e. $n=1$) as we shall investigate in detail, modified HDE density $\rho_{de}\propto L^{-1}$.

In DGP brane world model \cite{G. Dvali et al.,C. Deffayet}, our universe is considered as a flat, homogeneous and isotropic $3$-dimensional brane embedded in $5$-dimensional Minkowski bulk. And the Friedmann equation on the brane reads
\begin{equation}
\label{bfe}
H^2=\left(\sqrt{\frac{\rho}{3M^2_p}+\frac{1}{4r_c^2}}+\epsilon
\frac{1}{2r_c}\right)^2,
\end{equation}
 where $H=\dot{a}/a$ is the Hubble parameter, $a$ is the scale factor of the universe and $r_c$ is called crossover distance. For $r\ll r_c$
the DGP model degenerate to the usual $4$-dimensional Einstein's theory of  gravity. The effect of
high dimensional gravity emerge when $r\gtrsim r_c$. $\epsilon\equiv\pm1$ represents two branches of the model of which the $\epsilon=+1$ branch is the self-accelerating solution where the universe may accelerate in the late time purely due to the modification of gravity, while the expansion of the $\epsilon=-1$ branch is not able to speed up without dark energy.  In the model we investigate in this work, the energy density $\rho$ in Eq.(\ref{bfe}) contains both dust matter and dark energy, that is,  $\rho=\rho_m+\rho_{de}$ where $\rho_m=\rho_{m0}a^{-3}$ is the same as that in standard cosmology.

\section{Hubble horizon as IR cutoff}
\label{sec3}

It is natural to choose Hubble horizon $H^{-1}$ as the IR cutoff, and then $\rho_{de} \propto H$.
Defining that $\Omega_m=\frac{\rho_m}{3M_p^2H_0^2}=\Omega_{m0}a^{-3}$,
$\Omega_{r_c}=\frac{1}{4r_c^2H_0^2}$, $\Omega_{de}=\frac{\rho_{de}}{3M_p^2H_0^2}=AH$
and noting that Eq.(\ref{bfe}) can be equivalently written as
\begin{equation}
H^2-\epsilon \frac{1}{r_c}=\frac{1}{3M_p^2}\rho,
\end{equation}
 we obtain that
\begin{equation}\label{hh}
\left(\frac{H}{H_0}\right)^2-\left(2\epsilon\sqrt{\Omega_{r_c}}+AH_0\right)\left(\frac{H}{H_0}\right)=\Omega_m.
\end{equation}
Note that when $AH_0\rightarrow 0$, the original DGP brane world model is recovered. For the $\epsilon=+1$ branch,
Eq.(\ref{hh}) show that the model we considered here is nothing but a $\epsilon=+1$ branch in pure DGP model with a decreased crossover distance $\tilde{r}_c= {r_c}\left(1+AH_0^2r_c\right)^{-1}$.   For the  $\epsilon=-1$ branch, if
$AH_0^2r_c < 1$, the resulting model is still a $\epsilon=-1$ in pure DGP model in which the crossover is modified to be $\tilde{r}_c={r_c}\left(1-AH_0^2r_c\right)^{-1}$. However, if $AH_0^2r_c > 1$, the resulting model is equivalent to a $\epsilon=+1$ branch in pure DGP model with crossover distance $\tilde{r}_c={r_c}\left(AH_0^2r_c-1\right)^{-1}$. That is to say, the Modified HDE can drive the $\epsilon=-1$ branch accelerating.  Note that when $AH_0^2r_c$ is exactly equal to unity, the resulting model is reduced to be standard Einstein-de Sitter model.

\section{The future event horizon as IR cutoff}
\label{sec4}
The future event horizon is another choice for the IR cutoff. In fact, in the original version of HDE model \cite{Limiao}, the IR cutoff is just chosen to be the future event horizon of the universe, because Hsu \cite{Hsu} pointed out that if $L$ is taken to be Hubble radius, then it would yield a wrong equation of state (EoS) for dark energy.
When $L$ is taken to be future event horizon, we have
\begin{equation}\label{fh}L=a(t)\int^\infty_t\frac{dt'}{a(t')}=a\int^\infty_a\frac{da'}{Ha'^2}=\frac{A}{\Omega_{de}}.\end{equation}
On the other hand, Eq.(\ref{bfe}) can be rewritten as
\begin{equation}\label{rr}\frac{H}{H_0}=\sqrt{\Omega_m+\Omega_{de}+\Omega_{r_c}}+\epsilon\sqrt{\Omega_{r_c}}.\end{equation}
Therefore, we obtain the following integral equation
\begin{equation}\label{9}\int^\infty_a\frac{da'}{\left(\sqrt{\Omega_m+\Omega_{de}+\Omega_{r_c}}+\epsilon\sqrt{\Omega_{r_c}}\right)a'^2}=\frac{AH_0}{a\Omega_{de}},\end{equation}
and its differential form
\begin{equation}\label{10}
\Omega_{de}'=\frac{\Omega_{de}}{(1+z)}\left(1-\frac{\Omega_{de}}{AH_0\left(\sqrt{\Omega_m+\Omega_{de}+\Omega_{r_c}}+\epsilon\sqrt{\Omega_{r_c}}\right)}\right),
\end{equation}
where the prime denotes the derivative with respect to redshift $z$ which has the relationship with scale factor that $1+z=a^{-1}$ and $z=0$ at present. Eq.(\ref{10}) can be solved numerically and the initial condition can be set by
$\Omega_{de0}=1-2\epsilon\sqrt{\Omega_{r_c}}-\Omega_{m0}$, where $\Omega_{de0}$ denotes the present value of $\Omega_{de}$.

From conservation equation for the modified HDE, $\dot{\rho}_{de}+3H(\rho_{de}+p_{de})=0$,
EoS for the modified HDE $w_{de}\equiv p_{de}/\rho_{de}$ can be determined by
\begin{equation}\label{14}
w_{de}=-1+(1+z)\frac{\Omega_{de}'}{3\Omega_{de}}.
\end{equation}
Inserting Eq.(\ref{10}) into Eq.(\ref{14}),  we obtain that
\begin{equation}w_{de}=-\frac{2}{3}-\frac{\Omega_{de}}{3AH_0\left(\sqrt{\Omega_m+\Omega_{de}+\Omega_{r_c}}+\epsilon\sqrt{\Omega_{r_c}}\right)}.\end{equation}
Because $\Omega_{de}$ is always greater than zero, in both branches, the EoS of modified HDE is always less than $-2/3$.
Squaring  the both sides of Eq.(\ref{rr})
and comparing it with the Friedmann equation in $4$-dimensional standard cosmology
$\left(\frac{H}{H_0}\right)^2=\Omega_m+\Omega_{eff}$, we  obtain the effective dark
energy
\begin{equation}\label{Omega_eff}
\Omega_{eff}=\Omega_{de}+2\epsilon\sqrt{\Omega_m+\Omega_{de}+\Omega_{r_c}}\sqrt{\Omega_{r_c}}
\end{equation}
and its equation of state is determined by
\begin{equation}\label{w_eff}
w_{eff}=-1+(1+z)\frac{\Omega'_{eff}}{3\Omega_{eff}}.
\end{equation}
The deceleration parameter is
\begin{equation}q=-\frac{\ddot{a}}{aH^2}=\frac{H'}{H}(1+z)-1.\end{equation}
To illustrate the modified HDE  in DGP brane world, we plot the evolution of EoS of HDE $w_{de}(z)$, EoS of the effective dark energy $w_{eff}(z)$ and deceleration parameter $q(z)$ with respect to redshift $z$ in four figures.

\begin{figure}[!htbp]\begin{center}
\includegraphics[height=3cm]{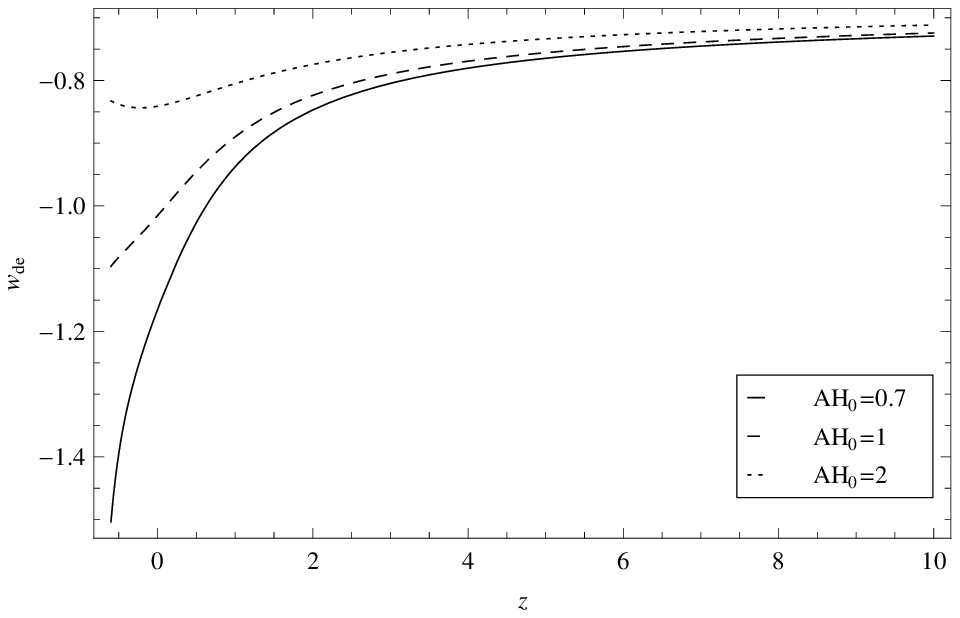}
\includegraphics[height=3cm]{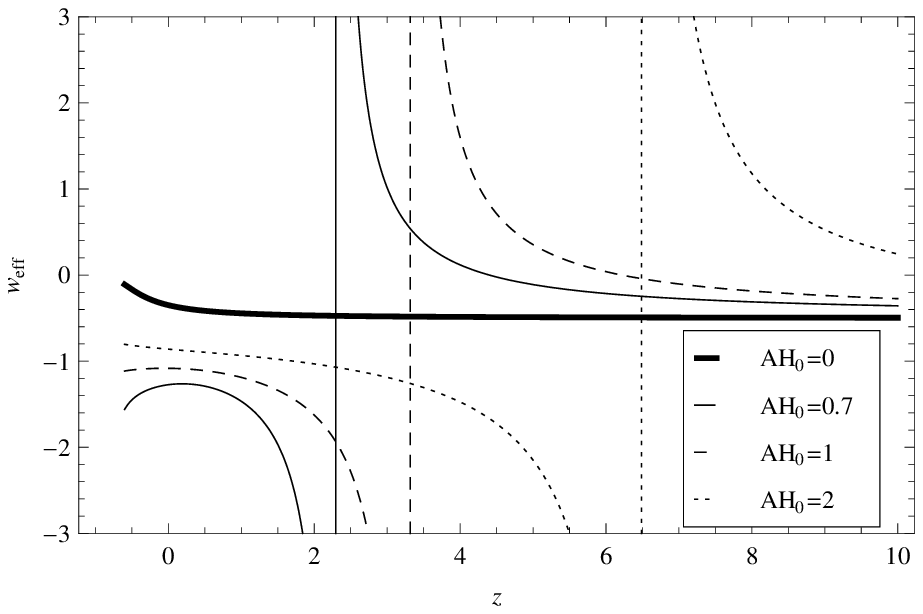}
\includegraphics[height=3cm]{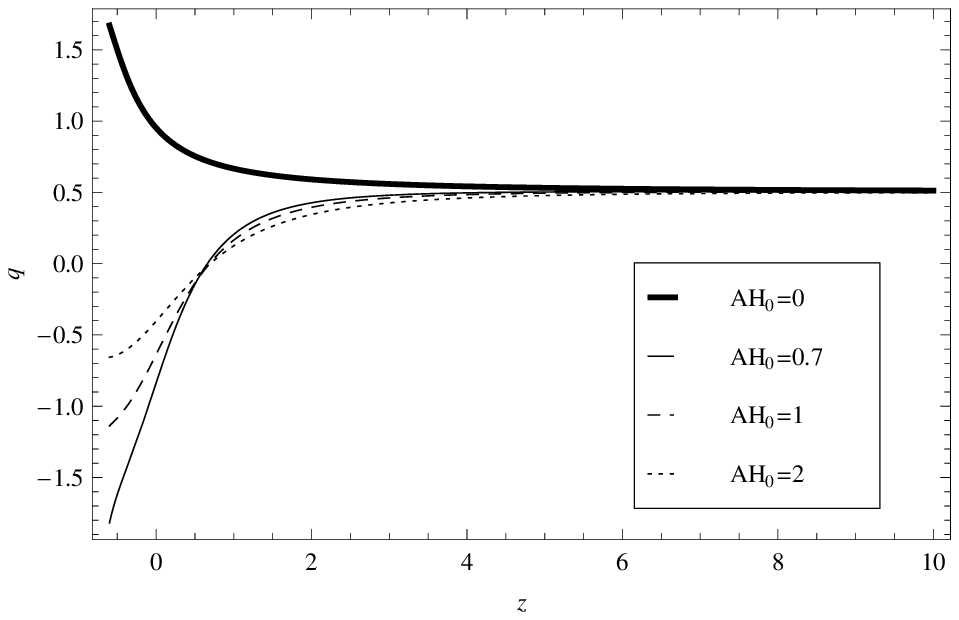}
\caption{The cosmological evolution of $w_{de}$, $w_{eff}$ and $q$ with
redshift $z$ in $\epsilon=-1$ branch when the future event horizon is taken as IF cutoff. Here we set $\Omega_{r_c}=0.03$ and $\Omega_{m0}=0.3$. }
\label{w_lamda}\end{center}\end{figure}

\begin{figure}[!htbp]\begin{center}
\includegraphics[height=3cm]{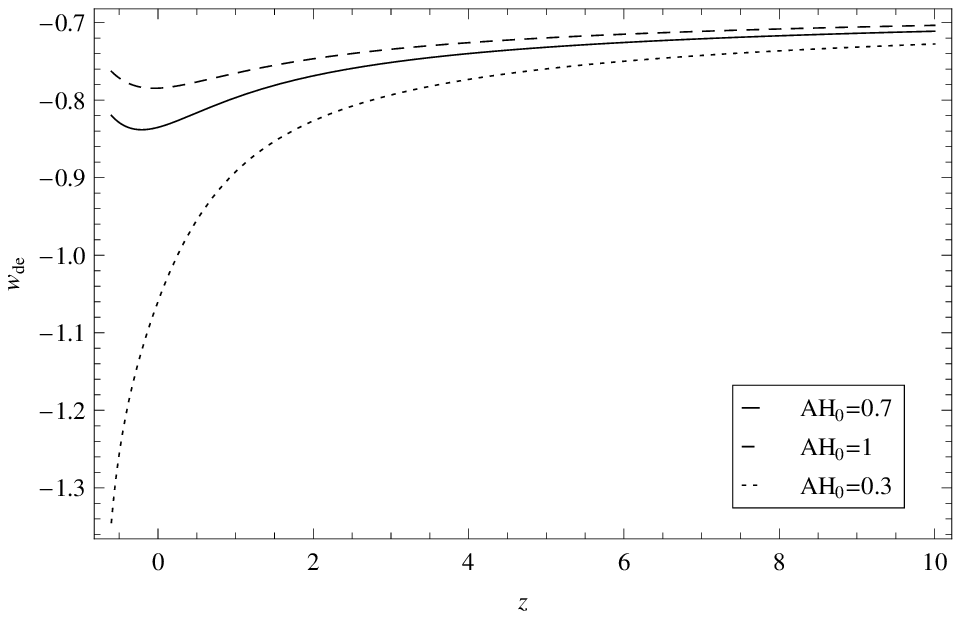}
\includegraphics[height=3cm]{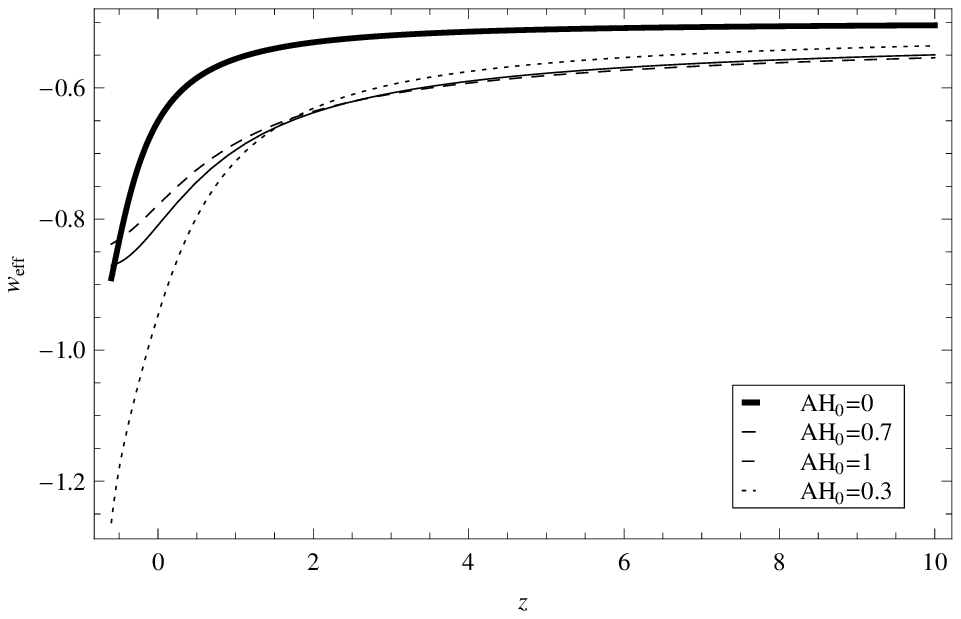}
\includegraphics[height=3cm]{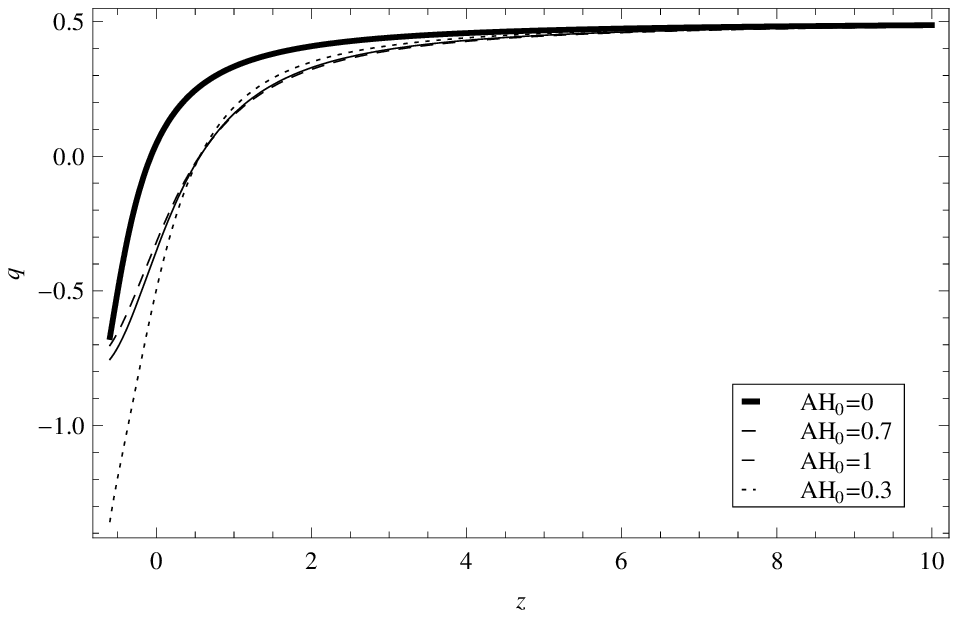}
\caption{ The cosmological evolution of $w_{de}$, $w_{eff}$ and $q$ with
redshift $z$ in $\epsilon=+1$ branch when the future event horizon is taken as IF cutoff. Here we set $\Omega_{r_c}=0.03$ and $\Omega_{m0}=0.3$. }
\label{w_lamda1}
\end{center}
\end{figure}

In Fig.\ref{w_lamda} and Fig.\ref{w_lamda1}, we set a given value of $\Omega_{r_c}$ and
choose different values of $AH_0$. Note that when $AH_0=0$, our model
 reduces to the pure DGP model.  To make a comparison with pure DGP brane model, we also plot the EoS of effective dark energy and deceleration parameter in the middle and right panels under the condition when $AH_0=0$.
From these two figures, we see that, in both $\epsilon=\pm 1$ branches, $w_{de}$ and $w_{eff}$ can cross phantom divide $w=-1$ from quintessence region $w>-1$ to phantom region $w<-1$ for some suitable parameters. It is found that both the $\epsilon=\pm1$ branches can undergo  an acceleration phase at late time, which is different from the pure DGP model, in which  only the $\epsilon=+1$ branch can accelerate. From Fig.\ref{w_lamda} we see also that $w_{eff}$ may become divergent for some value of $z$,  while the deceleration factor $q$ behaves well. It should be pointed out that the reason why $w_{eff}$ is divergent is that  the effective energy component $\Omega_{eff}$ expressed by Eq.(\ref{Omega_eff}) vanishes  for some value of redshift $z$ in $\epsilon=-1$ branch (while in $\epsilon=1$ branch, this situation will never happen), this leads to the failure of expression (17) for the value of $z$. However, this does not indicate  the $\epsilon=-1$ branch is not consistent with the evolution of the universe, because scale factor $a$, EoS of the modified HDE $w_{de}$ and Hubble parameter $H$ evolve smoothly.

\begin{figure}[!htbp]
\begin{center}
\includegraphics[height=3cm]{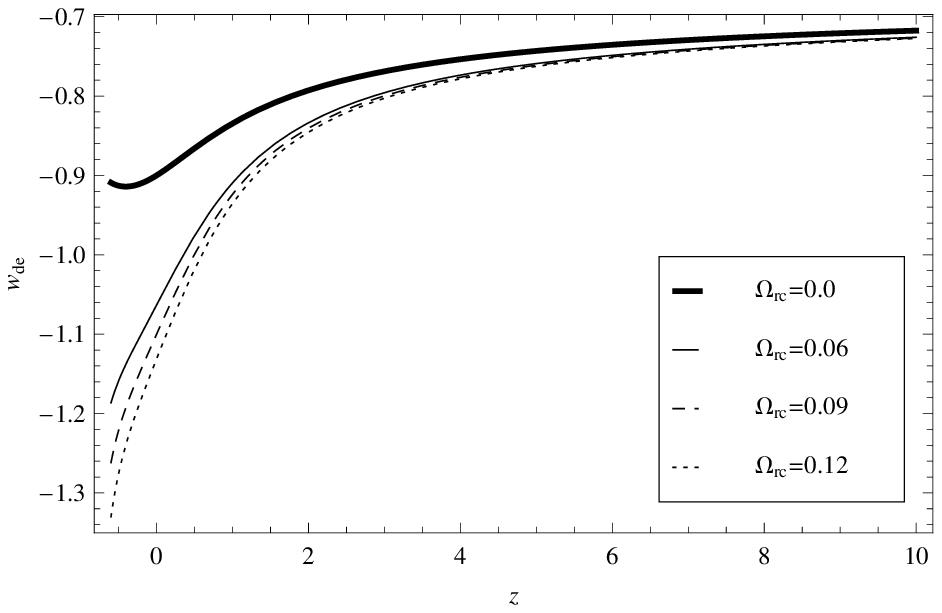}
\includegraphics[height=3cm]{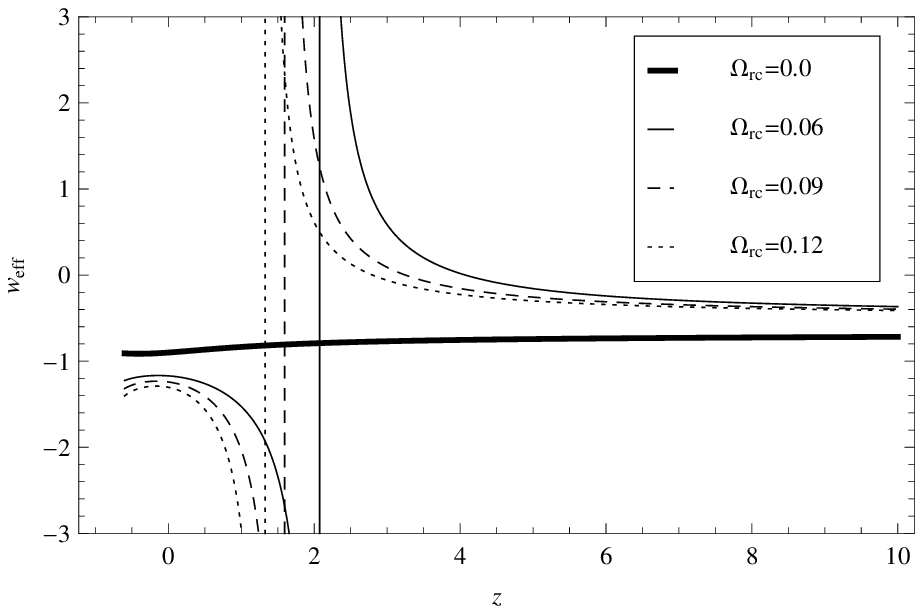}
\includegraphics[height=3cm]{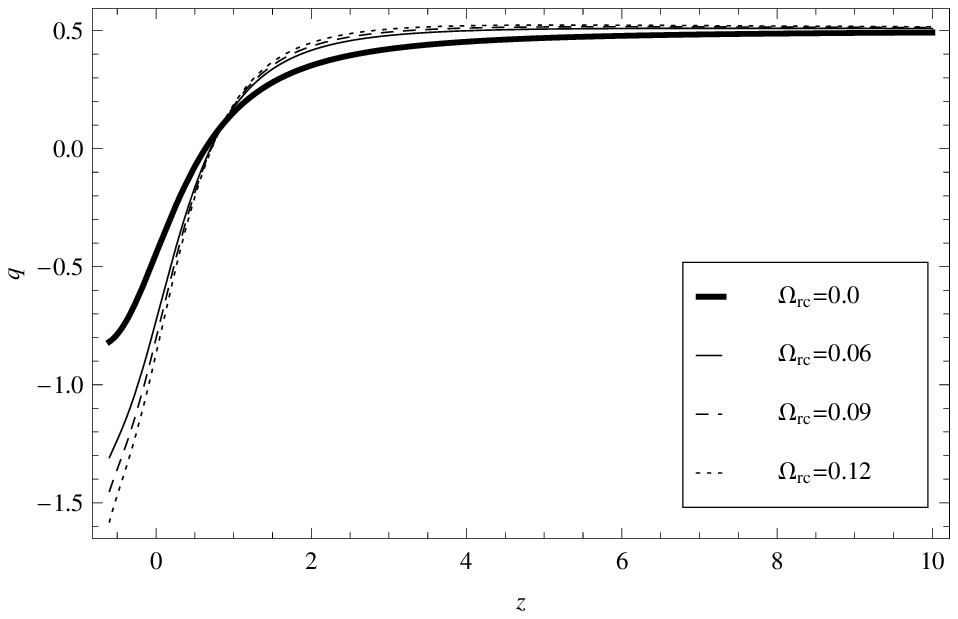}
\caption{The cosmological evolution of $w_{de}$, $w_{eff}$ and $q$ with
redshift $z$ in $\epsilon=-1$ branch when the future event horizon is taken as IF cutoff. Here we set $AH_0=1$ and $\Omega_{m0}=0.3$.}
\label{w_lamdaa}\end{center}\end{figure}

\begin{figure}[!htbp]
\begin{center}
\includegraphics[height=3cm]{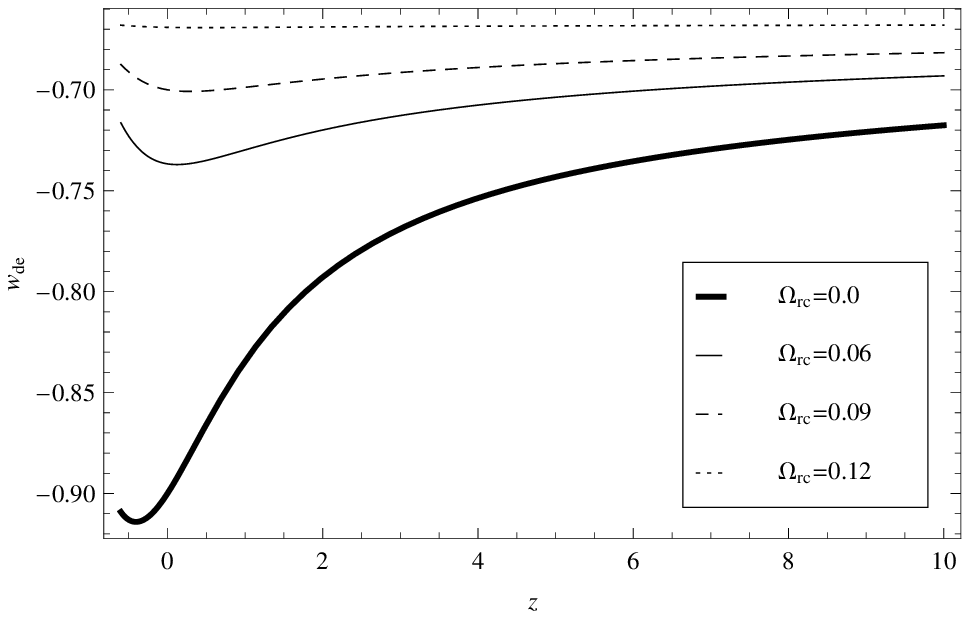}
\includegraphics[height=3cm]{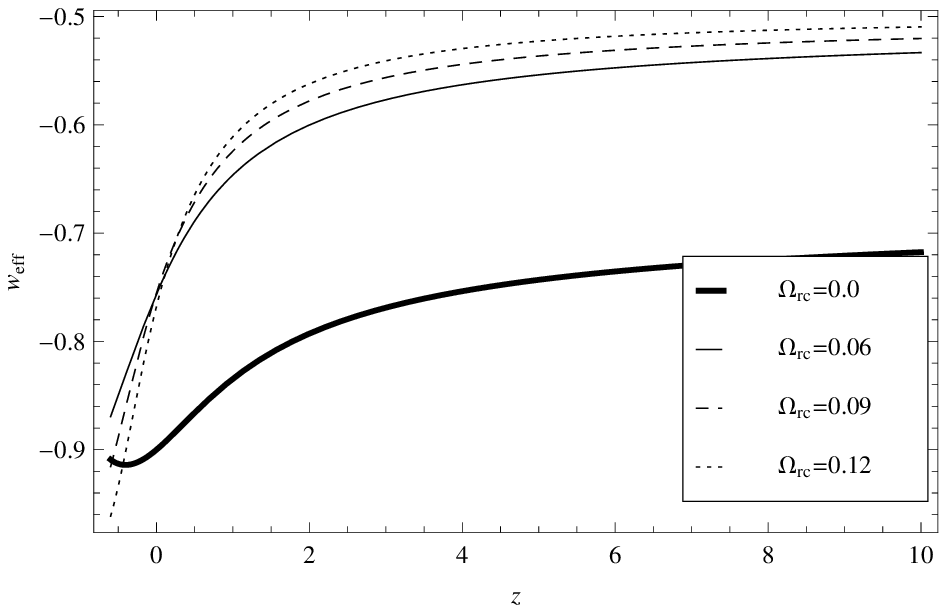}
\includegraphics[height=3cm]{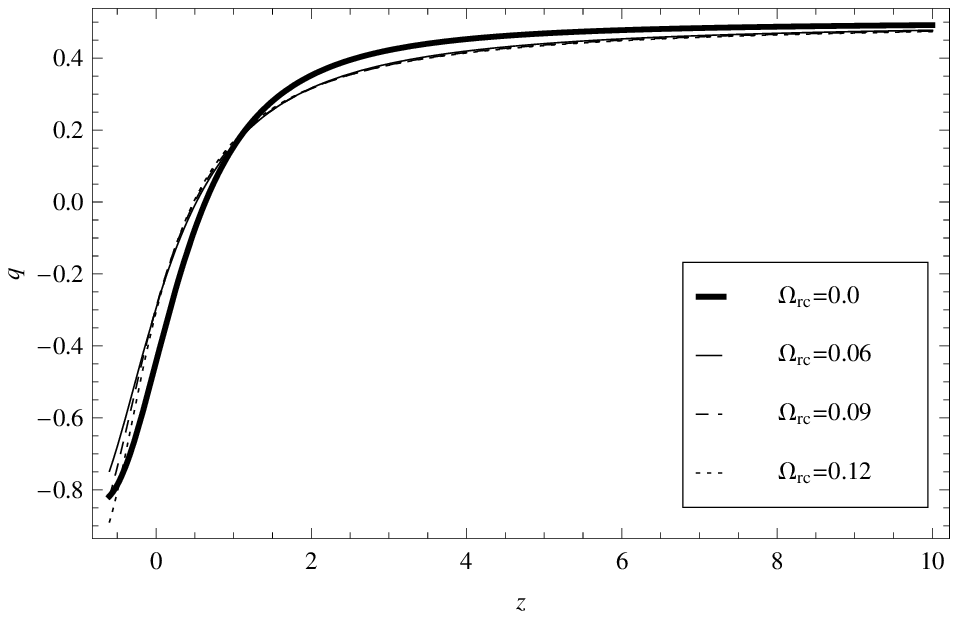}
\caption{The cosmological evolution of $w_{de}$, $w_{eff}$ and $q$ with
redshift $z$ in $\epsilon=+1$ branch when the future event horizon is taken as IF cutoff. Here we set $AH_0=1$ and $\Omega_{m0}=0.3$.}
\label{w_lamdab}\end{center}\end{figure}

In Fig.\ref{w_lamdaa} and Fig.\ref{w_lamdab}, we set a given value of $AH_0$  and
choose different values of $\Omega_{r_c}$. It is shown that one effect of $\Omega_{r_c}$ is to reduce the value of $w_{de}$ at late time in $\epsilon=-1$ branch, and is to increase $w_{de}$ value in $\epsilon=+1$ branch. However,
the decrease parameter is not heavily affected by the different value of $\Omega_{r_c}$ in both branches. This is not surprising because $\Omega_{r_c}$ in $\epsilon=-1$ branch hinders the expansion of the universe, while it drives the universe expanding in $\epsilon=+1$ branch. Therefore, the two effects of $\Omega_{r_c}$ counteract each other.

\section{Conclusions}
In all, we have investigated the cosmological dynamics of a modified holographic dark energy, which is derived from the UV/IR duality by considering the black hole mass in higher dimensions as UV cutoff, in DGP brane world model. We choose Hubble horizon and future event horizon as IR cutoff respectively. It is found that when Hubble horizon is considered as IR cutoff, the modified HDE behaves like an effect dark energy that modification of gravity in pure DGP brane world model acts and it can drive the expansion of the universe speed up at late time in $\epsilon=-1$ branch which in pure DGP model can not undergo an accelerating phase. When future event horizon acts as IR cutoff, it is shown that EoS of the modified HDE  can cross the phantom divide from quintessence region to phantom region during the evolution when a set of suitable parameters is chosen.  By considering the combination of the modified HDE and the $5$-dimensional gravity effect, we find the equation of state of the effective dark energy may also cross $w=-1$ divide, which may
lead to the Big Rip singularity.  Noting that the model we considered here is not the same as those in Refs.\cite{Saridakis} and \cite{Xing Wu}. In Ref.{\cite{Saridakis}, the author apply the bulk holographic dark energy in general $5D$ two-brane models and what the authors of Ref.\cite{Xing Wu} considered is  the evolution of the usual holographic dark energy in the DGP model.  

It is also worth pointing out that the future event horizon exists if and only if the universe is accelerating. Therefore, choosing future event horizon as IR cutoff will face circular reasoning or causality problem, see, for example, Refs.\cite{Gong:2010,cai} for detail comments on this aspect. However, from phenomenological and observational point of view \cite{gong2004}, it is still interesting to choose the future event horizon as IR cutoff.

\section*{Acknowledgments}
We would like to thank Chao-Jun Feng and Ying-Li Zhang for helpful discussions. This work is supported in part by National Natural Science Foundation of China under Grant  No. 10503002, Shanghai Commission of Science and technology under Grant No. 06QA14039 and Innovation Program of Shanghai Municipal Education Commission under Grant No. 09YZ148.


\end{document}